%% file: paper.tex
\documentclass[a4paper, 11pt]{article}
\usepackage{epsfig}

\newcommand{\be}{\begin{equation}}
\newcommand{\ee}{\end{equation}}
\newcommand{\ba}{\begin{eqnarray}}
\newcommand{\ea}{\end{eqnarray}}
\newcommand{\bi}{\begin{itemize}}
\newcommand{\ei}{\end{itemize}}

\newcommand{\nn}{\nonumber \\}

\newcommand{\Lp}{L_\perp} 
 
\newcommand{\tsig}{\tilde\sigma} 
\newcommand{\<}{\langle} 
\renewcommand{\>}{\rangle} 
\newcommand{\eq}{Eq.~}

\newcommand{\la}{\label}

\newcommand{\sun}{SU($N$) }

\begin{document}
\begin{titlepage}
\begin{flushright}
DESY preprint 05-069\\
\end{flushright}
\begin{centering}
\vfill

 \vspace*{2.0cm}
{\bf \Large Vortices on the worldsheet of the QCD string}

\vspace{2.0cm}
{\large Harvey~B.~Meyer}
\centerline{\sl DESY-Zeuthen,}
\centerline{\sl Platanen Allee 6}
\centerline{\sl D-15738 Zeuthen}
\vspace{0.1cm}\\
\centerline{\sl harvey.meyer@desy.de}

\vspace*{3.0cm}

\end{centering}

\centerline{\bf Abstract}
\vspace{0.1cm}
\noindent We investigate the properties of the QCD string in
the Euclidean SU($N$) pure gauge theory when the
space-time dimensions transverse to it are periodic.
From the point of view of an effective string theory, 
the string tension $\sigma$ and the low-energy constants $c_k$
of the theory are arbitrary functions of 
the sizes of the transverse dimensions
${\Lp}$.  Since the gauge theory is linearly confining
in $D=2,~3$ and 4 dimensions, 
we propose
an effective string action for the flux-tube energy levels 
at any choice of ${\Lp}$,
given $\sigma(\Lp)$ and $c_k(\Lp)$.
The L\"uscher term only 
depends on the number of massless bosonic degrees of freedom 
and the effective theory can account for
its evolution as a function of ${\Lp}$. 
As the size of one transverse dimension is varied,
we predict a Kosterlitz-Thouless transition 
of the worldsheet field theory at $\sigma(\Lp)\Lp^2\simeq1/8\pi$ 
driven by vortices, 
after which the periodic component of the worldsheet
displacement vector develops a mass gap and the 
effective central charge drops by one unit.
The universal  properties of the transition are emphasised.
\noindent 
\vfill

\vspace*{1cm}
\vfill
\end{titlepage}

\setcounter{footnote}{0}

\section{Introduction\la{sec:intro}}
Consider the Euclidean \sun pure gauge theory on a four-dimensional 
hypertorus of size $(L_1,L_2,L_3,L_4)$, with 
a pair of Polyakov loops running along the direction 4.
$L_3$ and $L_4\equiv L$ are fixed but very large compared
to all other length scales in the setup.
Direction 4 is interpreted as the time direction, so that the potential
between two static sources can be extracted from the Polyakov 
loop correlator:
\be
\< P_4(0)P_4^*(R\hat e_3)\>\equiv  e^{-V(R;L_1,L_2)L},\quad
L\gg R.\\
\ee

We are now interested in the system as a function of the size 
of the two other dimensions, $L_1$ and $L_2$. 
If one of them is kept large while the other is shrunk to zero, 
we will observe the `deconfining phase transition', 
where the centre symmetry along the finite dimension breaks spontaneously;
decreasing it further, eventually the theory dimensionally reduces to
a three-dimensional \sun gauge theory. 
If we now follow the same evolution, gradually reducing another
dimension, we again go through a deconfining phase transition,
followed by  a dimensional reduction to a two-dimensional theory.
At this point, the pair of Polyakov considered above
loops still obeys an area law; the theory is now perturbatively solvable. 

In fact, at any stage of this two-step dimensional reduction, 
the Polyakov loop correlator obeys an area law 
characterised by a string tension $\sigma(L_1,L_2)$ varying 
with the size of the transverse dimensions:
\ba
V(R;L_1,L_2)&\sim&\sigma(L_1,L_2)R, \qquad \sigma(L_1,L_2)R^2\gg 1.
\ea
An effective low-energy theory 
was recently proposed by L\"uscher and Weisz~\cite{lw04}
to describe the corrections to the classical string behaviour
$V(R)=\sigma R$, for the case of infinite transverse dimensions
in the three- and  four-dimensional SU($N$) gauge theories.
The central idea is to equate the Polyakov
loop correlator with the partition function of the fluctuating
$L\times R$ surface, corresponding to the string worldsheet.
The leading correction to the ground state energy\footnote{ 
In this paper we shall be interested only in the 
ground state of the quark-antiquark system, whose contribution
dominates the correlator  at large enough $L$.}
comes from Gaussian fluctuations:
\be
V(R)=\sigma R -\frac{\pi d_\perp}{24R}+O\left(\frac{1}{\sigma R^3}\right),
\ee
where $d_\perp$ is the number of (infinite)
transverse space-time dimensions.
It is the famous L\"uscher correction~\cite{luscher81}, 
in favour of which a significant amount of numerical evidence 
has been gathered~\cite{lw02,kuti}.

Since we have a linearly rising potential at any choice of $(L_1,L_2)$,
we can attempt to write an effective string theory for every 
$(L_1,L_2)$. As the latter are varied, we expect the effective
theory to interpolate between the purely four-dimensional,
the three-dimensional  and the (trivial) two-dimensional case.
The unknown parameters of the effective string theory now
become functions of $(L_1,L_2)$.  

When the transverse dimensions are large,
$\sigma$ becomes independent of their size.
As $L_1$  shrinks to zero,
$\sigma \propto 1/L_1^2$ (by asymptotic freedom); when $L_2\to 0$ as well,
$\sigma$ is proportional to $1/(L_1L_2)$. We note that
to discuss a smooth evolution of $V(R)$ as a function of $(L_1,L_2)$,
we must measure $R$ and $V(R)$ at each $(L_1,L_2)$ in units of 
$\sigma(L_1,L_2)$. From the point of view of the four-dimensional
theory, the increase of the latter as the transverse volume decreases
can be interpreted in terms of the overlap around the transverse
dimensions of the chromoelectric flux lines.
Assuming that the flux-tube energy density falls off transversely 
as $e^{-m_Gr_\perp}$, where $m_G$ is the scalar glueball mass
(something that is hard to establish numerically~\cite{sommer-bali-michael}), 
the corrections on $\sigma(L_1,L_2)$ would be $O(e^{-m_GL_{\rm min}})$, 
when $L_{\rm min}={\rm min}(L_1,L_2)$ is large.

In addition, there are  the `low-energy constants' 
$c_k$ appearing in the effective action describing the energetics of 
the worldsheet fluctuations beyond the quadratic order.
These functions of $(L_1,L_2)$ have to be given as input in
the effective theory. The only limit in which they are known
is the $L_1,L_2\to0$ limit, where they vanish.

However, the L\"uscher correction is completely independent of both 
$\sigma$ and the $c_k$.
The question that motivates the present paper is then the following. 
When the transverse dimensions $L_1,L_2$ are  large, the
coefficient of the L\"uscher correction for a bosonic string is
$-\pi/12$; when a transverse dimension shrinks to zero, 
the gauge theory should dimensionally reduce to a 
three-dimensional one, so the L\"uscher coefficient 
is then only $-\pi/24$.
The $1/R$ correction disappears altogether when a second dimension
is shrunk to zero. How and at what temperature do these transitions occur?
From the point of view of the effective string theory,
the coefficient is nothing but the central charge of the 
2d-conformal theory describing the fluctuations of the worldsheet;
it effectively counts the number of massless bosonic 
degrees of freedom living on the worldsheet.
\emph{We thus expect
the L\"uscher correction to have a discontinuity 
at a phase transition of the worldsheet quantum field theory.}

The component along a compact direction (say that of size $L_1$, while
$L_2$ stays large) 
of the worldsheet field $h(z_0,z_1)$ that describes 
its fluctuations becomes a \emph{periodic variable}. 
Since the worldsheet field theory is treated as an effective theory
with a fixed cutoff, it can be described as living on a lattice
with a fixed lattice spacing of order $\sigma^{-1/2}$. 
One is then led to consider, as effective partition function,
the class of statistical mechanics models which share the main features 
of the famous XY-model~\cite{kt}. 
In particular the vibrational modes of the 
string are expected to undergo a phase transition
of the  Kosterlitz-Thouless (KT) type~\cite{kt} related
to the proliferation of vortices on the string worldsheet. 
They have the effect of disordering the fluctuations by generating
a finite screening length. These vortices
are interpreted as tunnelling amplitudes between 
classical configurations of the string with different winding numbers.
The  KT criterion, based on an energy vs. entropy argument, yields
an estimate of the gauge-theory temperature where the transition occurs
(see section 4),
\be
T^* \equiv 1/L_1^*\simeq \sqrt{\frac{\sigma}{8\pi}}~.
\la{eq:T*}
\ee
The precise value of this temperature is non-universal; in particular,
the string tension appearing in the formula is $\sigma(L_1,L_2)$
and differs by a factor of order
one from the string tension at $L_1=\infty$.
Since the high-temperature phase of the XY model is dominated by 
vortex configurations and admits a finite mass gap (even in the infinite
worldsheet-volume limit), this transition naturally explains the disappearance
of the universal string correction associated with the periodic component of 
the worldsheet field $h$.
One of the aims of this paper is to describe the signatures of such a
Kosterlitz-Thouless transition in the spectrum of the QCD flux-tube.
In this respect, one might worry that $T^*$
is parametrically of order one in units of the string tension,
so that, in the four-dimensional language,
there is no parametric suppression of the string internal
degrees of freedom at that temperature. 
But these effects are described in our framework through
the dependence of $\sigma$ and the $c_k$ on the size 
of the transverse dimension.

\paragraph{}
The rest of this paper is organised as follows. In section 2
we review the properties of the 2D XY model relevant for our 
discussion of the effective worldsheet theory. 
The notation for the latter are set in section 3, 
and section 4 describes the main results on 
the effect of a finite periodic transverse dimension. 
A summary and outlook follows.
\section{On the Kosterlitz-Thouless
         phase transition\la{sec:KT}}
We review some well-known facts about the XY-model and
the Kosterlitz-Thouless phase transition. For more details,
we refer the reader to the abundant literature on the 
subject~\cite{kogut,zinn,drouffe,cardy}.

We take as starting point the $O(2)$ non-linear 
$\sigma$-model. Once the 2-component field is parametrised
as $(\cos\theta,\sin\theta)$, the continuum action is 
deceptively simple:
\be
S=\frac{\kappa}{2}\int d^2x \partial_\mu \theta \partial_\mu \theta.
\la{eq:O2}
\ee
What makes the dynamics non-trivial is that
$\theta$ is an angle variable. 
Therefore it cannot be rescaled, and $\kappa$ is not a redundant variable.
The (IR-regularised) two-point function in the Gaussian approximation
is thus given by  $-\frac{1}{2\pi \kappa}\log\left(\frac{r}{r_o}\right)$.
Such a perturbative analysis
around the trivial vacuum $\theta\equiv0$ of course does not
take the periodicity into account, and therefore the 
dynamics is that of an ordinary 
massless scalar field in that approximation.
A convenient way to deal with the periodicity of $\theta$
is to let it take values in {\bf R} but identify points
differing by $2\pi$. This means that a configuration of 
$\theta$ can be multivalued, as long as $e^{i\theta}$ is not.

The classical equation of motion $\Delta \theta=0$,
together with the boundary condition that $\theta$ goes to a constant
at infinity, implies that either $\theta$ is constant or has
singularities. 
If the latter are isolated, the most general solution is~\cite{zinn}
\be
e^{i\theta(z)}=e^{i\theta_\infty}\prod_{i=1}^{n}
\frac{(z-z_i^+)/|z-z_i^+|}{(z-z_i^-)/|z-z_i^-| }\la{eq:vort}
\ee
The interpretation of the solution is that \emph{vortices} of intensity
$\pm1$ are centred at the points $z_i^\pm$. Higher intensities
are obtained when several $z_i^+$ (or several $z_i^-$) coincide.
The angle $\theta$ is a multivalued function, but $e^{i\theta}$
is well-defined everywhere, except at the singular points.

Although the energy of vortex configurations is divergent both 
in the infrared and in the ultraviolet, 
in the (lattice) regularised theory their entropy 
is of the same functional 
form\footnote{$L$ is the linear size of the system;
the energy of a vortex-antivortex pair is roughly obtained
by replacing $L$ by their separation.} $\log L/a$.
So, depending on the temperature, they can
play an important role in the partition function. 
The estimate for the critical temperature one obtains from this
\emph{Kosterlitz-Thouless criterion} is 
\be
\kappa^*=\frac{2}{\pi}
\ee

These considerations show that to compute the corrections
to the free theory, we need to formulate it on a lattice
and generalise the perturbative approach by taking into 
account Gaussian fluctuations around not only the trivial 
vacuum, but also the vortex configurations:
\be
S[\theta=\theta_{\rm vort}+\delta\theta]=
S_{\rm vort}[\theta_{\rm vort}] + 
\frac{\kappa}{2}\int d^2x \left(\partial_\mu (\delta\theta)\right)^2
\ee
Since we have to sum over the different vortex configurations
$\theta_{\rm vort}$, the system
now appears to describe a gas of classical charged particles
interacting via a two-dimensional Coulomb potential and globally
neutral~\cite{kogut} (the neutrality constraint comes from the 
infrared divergence in the $\theta$ two-point function). 
From one's intuition of a Coulomb gas,
one infers that these charges are bound into dipoles 
by a two-dimensional Coulomb potential and thus
form a dielectric medium at low temperature. At high temperature,
the charges instead form a plasma which has a finite correlation 
length due to Debye screening.

The continuum action~(\ref{eq:O2}) thus has a line of fixed points,
$0>\kappa\geq \kappa^*$. 
To investigate the theory at all values of $\kappa$,
one may then choose a lattice action, which
has to preserve the essential periodicity property
of the continuum action and have a local minimum at $\theta=0$.
The most general form of the partition function is~\cite{cardy}
\be
Z = \int \prod_x \frac{d\theta(x)}{2\pi}
\prod_{(xy)}\sum_{m(x,y)=-\infty}^\infty  
e^{-J(m(x,y)) + im(x,y)(\theta(x)-\theta(y))}\la{eq:general}
\ee
where the sum is over near neighbours $(xy)$ 
The function $J$ must be even and have the asymptotic behaviour
$J(m)\propto m^2$, so that the Gaussian action~(\ref{eq:O2})
is recovered in the naive continuum limit.

The simplest choice is the so-called XY model~\cite{kt}:
\be
S_{\rm xy}=K_{\rm xy}\sum_{x,\mu}
\left[1-\cos(\theta(x)-\theta(x+a\hat e_\mu))\right].
\ee
This corresponds to $J(m)=-\log I_m(K_{\rm xy})$, 
where $I_m$ is the modified Bessel function.
At very small $K_{\rm xy}$ it is easy to see by strong coupling 
expansion~\cite{kogut} that the mass gap is given by $\log1/K_{\rm xy}$.
When the transition is approached from the high-temperature phase, 
the correlation length in lattice units diverges as
\be
\xi=\xi_o\exp\left(b\sqrt{\frac{K^*_{\rm xy}}{K_{\rm xy}^*-K_{\rm xy}}}\right),
\la{eq:xi}
\ee
where the functional form is universal, while 
$\xi_o$, $b$ and $K^*_{\rm xy}$ are not.
At very large  $K_{\rm xy}$, the naive continuum limit~(\ref{eq:O2})
is the correct one with $\kappa=K_{\rm xy}$. 

\paragraph{}
Another standard partition function, with $J(m)=\frac{m^2}{2K_{\rm v}}$,
is the Villain model~\cite{villain}: 
\ba
Z_{\rm v}&=&
\int \prod_x\frac{d\theta(x)}{2\pi} ~\prod_{x,\mu} 
z(\theta(x)-\theta(x+a\hat e_\mu)),\\
z(\theta)&=&\sqrt{2\pi K_{\rm v}}\sum_{n=-\infty}^\infty \exp\left\{
   -\frac{K_{\rm v}}{2}\left(\theta-2\pi n\right)^2\right\}.
\ea
Since the asymptotic behaviour of $J(m)$ in the XY-model is 
$m^2/2K_{\rm xy}$, at large $K_{\rm v}$ the Villain 
probability distribution becomes identical to the XY probability 
distribution with $K_{\rm v}=K_{\rm xy}$. 
Thus the continuum limit is again (\ref{eq:O2}) with
$\kappa = K_{\rm v}$ in that regime.
In general, the vortices decrease the effective $\kappa$
(defined as the $\kappa$ that would give the same two-point 
correlation function in the Gaussian approximation~\cite{drouffe}). 
If $R_o<a$ is the radius of the vortex-core, the relation is 
\be
\frac{1}{\kappa_{\rm eff}}=\frac{1}{K_{\rm v}}+
\frac{\pi^2}{2}\frac{1}{K_{\rm v}-2/\pi} 
\left(\frac{R_o}{a}\right)^{2\pi K_{\rm v}}.\la{eq:kappa_eff}
\ee
Thus the correction is exponentially small at large $K_{\rm v}$.

A duality transformation brings the Villain
partition function into the following
form, defined on a dual lattice:
\be
Z_v=\int \prod_x d\phi(x) \sum_{m(x)=-\infty}^\infty \exp\left\{
-\frac{1}{2K_{\rm v}}\sum_{x,\mu} \left( \phi(x)-\phi(x+a\hat e_\mu)\right)^2
+ 2\pi i \sum_r m(x)\phi(x) \right\},\la{eq:dualV}
\ee
where $m(x)$ is a dynamical variable taking integer values.
It can be shown~\cite{balog-priv-comm} 
to be equivalent to the (naively discretised) sine-Gordon model
\be
S_{\rm sG}=\sum_{x} \left[
\frac{1}{2}\sum_\mu\left(\phi(x)-\phi(x+a\hat e_\mu)\right)^2
 -\frac{m_o^2}{\beta^2}\cos\{\beta \phi(x)\} \right].
\ee
When the fugacity of vortices is small 
(this is universally the case on an RG trajectory close 
to the continuum for $\kappa>\kappa^*$),
one need keep only the terms $m(x)=0,\pm1$ in (\ref{eq:dualV})
and the bare lattice parameters are then related by~\cite{kogut}
\be
\beta = 2\pi \sqrt{K_{\rm v}},
\qquad m_o^2=8\pi^2 K_{\rm v}~  e^{-\pi^2K_{\rm v}/2}.\la{eq:beta}
\ee
What is relevant to our discussion is that
the continuum limit of the sine-Gordon model at 
\be
\beta^*=\sqrt{8\pi} \la{eq:beta*}
\ee
can be used to describe the  Kosterlitz-Thouless 
transition~\cite{amit,balog00}. 

The sine-Gordon model is exactly solvable in the continuum~\cite{zamolo} 
by bootstrap methods for $0\leq \beta^2 < 8\pi$, in which range 
the potential term is relevant~\cite{zinn}; $m_o$ renormalises but 
the coupling $\beta$ does not~\cite{zinn}. Remarkably, 
there is even an exactly solvable (light-cone lattice)
regularisation of the model~\cite{DdV}.
\section{The effective theory for the QCD string\la{sec:eff_th}}
We briefly remind the reader of the effective theory for the 
QCD string~\cite{lw04}. It is in principle capable of 
predicting the splittings (and the degeneracies)
between low-lying energy levels of the flux-tube, 
once a few low-energy constants have been determined. 
So it is a low-energy effective theory with a finite
UV-cutoff set by the string tension.
Indeed, at energies of that order, once expects the internal 
degrees of freedom of the string to become excited.

This theory is specified by a 
partition function associated with
an action living on the worldsheet $R\times L$ of
the string. The connection with the gauge theory
observables is that the Polyakov loop correlation function
$\< P_t(0)^*P_t(x)\>$ with $|\vec x|=R$ is proportional to 
this partition function.
The latter may be considered as the partition function
of a non-renormalisable Euclidean quantum field
theory in (1+1) dimensions, on a spatial `volume' 
of linear size $R$ and at temperature $T=1/L$;
or, alternatively, as a 
statistical mechanics system living on a
two-space-dimensional lattice with a lattice spacing of 
order $\sigma^{-1/2}$ (which has nothing to do with 
the lattice on which the gauge theory may 
or may not be regulated).  

The field living on the worldsheet $(0\leq z_0\leq L,
0\leq z_1\leq R)$ is a bosonic field $h$
with $D-2$ components. Since it represents the displacement 
of the world sheet from the classical configuration, it has 
engineering dimension of length.
The Polyakov loops are the propagators of static quarks,
so we are discussing the open-string case, and this is 
reflected in the effective theory by imposing Dirichlet 
boundary conditions on $h$ at $z_1=0$ and $z_1=R$.
The action of the effective theory is
\be
S=\sigma RL + \mu L + S_0 + S_1 + S^{(1)}_2 + S^{(2)}_2 + \dots
\ee
where the Gaussian action
\be
S_0 = \frac{\tilde\sigma}{2}\int d^2z ~~\partial_\mu h\cdot \partial_\mu h
\la{eq:gauss_action}
\ee
gives the leading contribution to the string spectrum at large $R$.
Each higher-order term is multiplied by an unknown dimensionless constant
$\tilde c_k^{(i)}$, times $\tilde\sigma$;
some of them vanish/are constrained by the open-closed 
string duality~\cite{lw04}. The lower index on the action terms 
above indicate how many more derivatives of $h$ they contain than
$S_0$.
The parameter $\tilde\sigma$ controls the amplitude of the fluctuations
around the classical solution: in this form, the whole action of
the effective theory is multiplied by it.
Of course, $\tilde\sigma$ is of the same order as $\sigma$ itself, 
and the fact that we are to treat the latter as large 
(it is the UV cutoff)
shows that we are in fact doing a semi-classical calculation.
But if we introduce the canonically normalised
field $\phi=\sqrt{\tilde\sigma}h$, $\tsig$ is absorbed into the low-energy 
constants, $c_k^{(i)}\equiv\tilde c_k^{(i)} /(\tilde\sigma)^{k/2}$.
The value of  $\tsig/\sigma$ is not observable here:
if for instance the low-lying energy levels are 
determined from lattice simulations, 
one will simply  determine the dimension-full
constants $c_k^{(i)}$ in units of $\sigma$.

It should be noted that for 
any relativistic bosonic string, one has in fact 
$\tilde\sigma=\sigma$: \emph{the amplitude of 
Gaussian fluctuations are dictated by the same string tension 
that determines the energy of the classical configuration.}
It appears however that one cannot test this property for the 
flux-tube by measuring the low-energy spectrum in infinite transverse
dimensions. From now
on we drop the tilde on $\sigma$, but it is good to keep the 
distinction in mind.
\section{The effect of a periodic transverse dimension}
We now consider the effective theory describing
the Polyakov loop correlator, for simplicity,
in $D=3$ dimensions -- we will later return to the $D=4$
situation. The transverse dimension
(the one in which the unique component 
of $h$ can fluctuate) is periodic, with period $L_1\equiv\Lp$: 
\be
\< P_3(0)P^*_3(R\hat e_2)\> \equiv e^{-V(R,\Lp)L}, 
\quad L_3\equiv L\gg R.
\ee

How is one to generalise the effective string theory
discussed in the previous section to include the effect
of the periodicity of the transverse dimension?

The general requirements on the effective action  
(described in detail in~\cite{luscher81})
that lead to the form~(\ref{eq:gauss_action}) must be kept,
but $h$ is now a periodic variable: $h + \Lp$ 
must be identified with $h$. It is then convenient
to rescale $h$ according to 
\be
h = \frac{\Lp\theta}{2\pi}.
\ee
Because of this periodicity, a convenient way to
write the most general partition function is to formulate
it on a lattice with lattice spacing of order $1/\sqrt{\sigma(\Lp)}$. 
Working on a lattice is equivalent to working with a
continuum effective action with a fixed momentum cutoff~\cite{symanzik}.
Our Ansatz for the effective partition function is
\ba
Z(\Lp,L,R)&=&e^{\sigma(\Lp)RT+\mu T} 
\int \prod_x \frac{d\theta(x)}{2\pi}
\prod_{(xy)}\sum_{m(x,y)\in Z } 
e^{-J(\Lp,m(x,y)) + im(x,y)(\theta(x)-\theta(y))}\nn
J(\Lp,m)&\sim &m^2/(2K),\quad m\to \infty.\la{eq:Z}
\ea
where $(xy)$ is a pair of nearest neighbours and
\be
K \equiv \frac{\sigma(\Lp)\Lp^2}{(2\pi )^2}.\la{eq:KK}
\ee
The naive continuum limit is then given by
\be
S_o=\frac{K}{2}\int d^2z~ \partial_\mu \theta ~\partial_\mu \theta~.
\la{eq:naive}
\ee
Interactions over a few lattice spacings are possible but should
not affect the universal properties of the system, as long as 
the naive continuum limit is the one given in~(\ref{eq:naive}).
The simple application of the 
Kosterlitz-Thouless analysis leads to the 
following scenario as a function of $\Lp$ 
(assuming $R,L$ very large):
\bi
\item for large $\Lp$ (low-temperature phase
of the XY model), the Gaussian approximation is good, 
and the periodicity of $\theta$
plays no role for the long-distance
properties of the theory. In particular, 
the mass gap vanishes and the universal part of the 
Casimir energy of the system remains exactly $\frac{-\pi}{24R}$.
\item as $\Lp$ decreases, the vortices 
can affect the non-universal contributions to the 
energy levels.
\item when $\Lp$ reaches $\Lp^*$,
vortices on the worldsheet become relevant 
and the Kosterlitz-Thouless transition takes place.
The KT estimate of $\Lp^*$ is given by
\be
   \Lp^* \simeq \sqrt{\frac{8\pi}{\sigma(\Lp)}}~.
\la{eq:Lp*KT}
\ee
\item for $\Lp<\Lp^*$, the theory has a finite mass gap $m$.
The Casimir energy if of order $e^{-2mL}$, and
the fluctuation along this dimension is effectively `switched off'. 
\item
When $\Lp\ll \Lp^*$, the strong coupling expansion (see e.g.~\cite{kogut})
tells us that to leading order in $\kappa$, $m=-\log\kappa$.
\ei
It can be shown~\cite{drouffe} that the KT transition takes place
at somewhat larger $\Lp$ than the KT estimate~(\ref{eq:Lp*KT}), 
because of the disordering effect of vortices.
$K^*$ 
can be extracted from lattice simulations: 
first the function $\sigma(\Lp)$ must be computed, and then 
$K^*$ can be obtained from the determination of $\Lp^*$.
It should be noted that it is a non-universal quantity; 
for the square-lattice XY-model, it was
found in numerical simulations~\cite{hasenbusch} that 
$\kappa^*=1.1199(1)$, relatively far from the Kosterlitz-Thouless
estimate of $2/\pi\simeq0.6366$. 

\paragraph{Physical interpretation of vortex configurations  }
It is useful to complement the machinery of two-dimensional 
statistical mechanics with some more intuitive 
considerations. The action~(\ref{eq:Z}) is a saddle-point
expansion around the classical string solution which
connects a quark to an antiquark separated by distance $R$.
Clearly, there are also classical solutions 
with a net winding number
around the compact transverse dimension. They have energy 
\be
E_{\rm cl}(n)= \sigma(\Lp) \sqrt{R^2+n^2\Lp^2}\simeq 
E(n=0)+ \frac{\sigma(\Lp) \Lp^2n^2}{2R}, \quad R\gg \Lp.
\ee
The $n=1$ classical solution becomes as light as the
first Gaussian excitation of the  $n=0$ solution when
$\sigma(\Lp) \Lp^2=2\pi$. The different classical configurations
can only be connected if the string goes through more energetic
configurations. Therefore such transitions are classically forbidden
tunnelling processes. Quantum mechanically, one expects the tunnelling
to become a frequent fluctuation when the zero-point energy 
of the $n=0$ vacuum is of same order as the gap $E(1)-E(0)$.
In this context, the worldsheet direction of size $R$ is 
interpreted as the space-direction, and that of size $L$ as
the Euclidean time direction in a path integral treatment
of the quantum-mechanical string.
A vortex being a point-like object on the worldsheet, 
it corresponds to a \emph{process} of the string describing
a transition from a state at $t=-\infty$ to another state
at $t=+\infty$. As long as we are considering the asymptotic
limit $L\to\infty$, only vortex-antivortex configurations
have a finite free energy in the 2d QFT. By writing the solution
(\ref{eq:vort}) with a single ${\rm v}\bar{\rm v}$ pair explicitly,
it is easy to see that a worldsheet containing a 
${\rm v}\bar{\rm v}$ pair describes the `life' of a string that had
winding number $n=0$ at $t=-\infty$, goes through a state with
winding number $\pm1$, and returns to $n=0$ classical state (see
Fig.~\ref{fig:winding}).

One may find the discussion analogous to the discussion of 
instantons in the 4D infinite volume gauge theory. The (anti)vortices 
correspond to (anti)instantons, the classical configurations of the 
string correspond to the configurations of the gauge field at fixed $t$
where the mapping from the sphere $S^2$ at spatial infinity 
to SU(2) has a definite winding number.
A difference is that, due to infrared divergences,
 in the 2d system only configurations satisfying
$\sum_m m(n_m-\bar n_m)=0$ are allowed, where $n_m$ ($\bar n_m$) 
is the number of (anti)vortices of strength $m$ in the configuration.
By contrast, there are configurations of the 4D gauge theory 
with a net topological charge which have a finite Euclidean action.

\input{pic1-insert.tex}
\paragraph{Effect of the vortices on the 
           string spectrum at  $\Lp>\Lp^*$ }
For $\Lp>\Lp^*$, we are in 
the low-temperature phase of the XY model, and vortices
correspond to an irrelevant operator. Therefore they do not
affect the universal $1/R$ correction, which is
determined by the canonical kinetic term, the only 
renormalisable one compatible of the symmetry requirements
on the effective action. 

More generally, $E_C(R)R$ and $E_1(R)R$, 
the Casimir energy and
the energy gap to the first string excitation,
are low-energy physical observables in the fixed-cutoff
worldsheet quantum field theory; in fact, the latter
quantity was used as a definition of the 
renormalised coupling in the XY-model in~\cite{wolff}.
According to Symanzik's theory
on the approach to the continuum limit of physical 
observables, irrelevant operators give a contribution
which is suppressed by the lattice spacing to the 
power of their scaling dimension minus the number 
of space-time dimensions~\cite{symanzik}. 
In particular the higher-derivative interactions
on the worldsheet give contributions to the energy levels
which are suppressed with respect to the $1/R$ term
by $1/\sqrt{\sigma(\Lp)}$ to the power
of the number of derivatives of the irrelevant operator,
minus two. Now as long as the theory is scale invariant,
dimensional analysis tells us that the power of $1/R$
that suppresses the contribution of an irrelevant 
operator to $E(R)\cdot R$ must be this same power.
For instance, $(\partial_\mu \phi)^2 (\partial_\nu \phi)^2$
contributes as $1/(\sigma(\Lp) R^2)$ to $E\cdot R$,
if $E$ is a string energy level.
Thus the parametric form of the energy levels computed 
in~\cite{lw04} could be guessed on general grounds,
while relating the coefficients to the action 
and computing the string degeneracies requires
an explicit calculation. Similarly the vortex-operator 
gives a suppressed contribution.
Its scaling dimension, as computed in the Gaussian theory 
$S=\frac{\kappa}{2}\int d^2x (\partial_\mu \theta)^2$,
is~\cite{cardy} 
\be
x_{\rm v}=\pi \kappa. \la{eq:xv}
\ee
Now, since the free energy of a single vortex is divergent
in the infrared, the single-vortex contribution to the partition
function vanishes in the thermodynamic limit of the 
2d system~\cite{cardy}. The first contribution comes from a
vortex---anti-vortex pair, so that the vortex contribution
to the string ground state energy is suppressed by $1/R$ to 
the power $2x_{\rm v}-2$:
\be
\Delta E(\Lp,R) = \frac{C}{R}~(\sigma(\Lp) R^2)^{1-\pi \kappa}
\ee
The peculiarity is that the power varies continuously
with the transverse size. 
We note that it can also be obtained by using the connection with
the sine-Gordon model and the fact that the scaling dimension
of the  $\cos{ \beta\phi(x)}$ operator is 
$\frac{\beta^2}{4\pi}$~\cite{zinn}.

We now discuss the relation between $\kappa$ and $\Lp$.
The lattice theory, at lattice spacing $a\equiv 1/\sqrt{\sigma(\Lp)}$, 
is initially parametrised by the coupling 
$K\equiv \sigma(\Lp) \Lp^2/4\pi^2$.
Under the renormalisation group the coupling $K$ evolves
as a function of $\mu a$, where $\mu$ is a renormalisation scale,
in order to keep the long-distance physics constant. 
Under coarse-graining, the theory flows into a Gaussian fixed point, with 
a Gaussian coupling $\kappa\equiv \lim_{\mu a \to \infty}K(\mu a)$, 
which is a non-universal function of the initial $K$.
All we can expect is that $\kappa$ is monotonously increasing in $K$. 

To determine the RG flow of $K$ in a given model, we have
to choose a physical quantity that we keep fixed 
as the lattice spacing is varied. One possibility is to
choose the ``spin-spin'' correlation function, 
$\< e^{i(\theta(0)-\theta(x))} \>$, which in the Gaussian
theory behaves as $C(a/|x|)^{1/2\pi \kappa}$ at large distances.
In the lattice theory at lattice spacing $a$, 
the long distance behaviour of this 
correlator will also decay algebraically, with a certain
exponent $b$. Then the renormalised $K(\mu a)$
can be set to $1/2\pi b$. In the Villain (also called `periodic Gaussian')
model, this computation
can be carried out explicitly (see section 2, Eq.~\ref{eq:kappa_eff}).
At large $K=K_{\rm v}$, the shift from $K$ to $K(\mu a)$
is exponentially small in $K$. In that specific model, the vortices
are the only origin of renormalisation. The exponential suppression
of their effect is a universal fact, because their fugacity is 
of order $y=e^{-c K}$, where $c$ is however a non-universal quantity
representing the core energy of a vortex. 
In general, there is in addition the more usual
renormalisation of $K$ due to
operators containing a higher number of derivatives. 
These effects are of order one.
So although the scaling dimension of the vortex operator
is a non-universal quantity, at large $K$, we have
\be
\kappa = K~+ O(1).
\ee

\paragraph{Properties of the worldsheet phase transition}
We now proceed to discuss
the universal signatures of the KT transition, 
from the point of view of string observables.

A first point to note is that the critical 
value of $\kappa^*=2/\pi$ is universal 
(while that of $K^*$ is not). Therefore
the vortex-antivortex pair contribution to
the string energy levels reaches  
\be
\Delta E\propto 1/\sigma(\Lp^*) R^3
\ee
at the KT transition. At the KT point, the fluctuation-field
correlations on the worldsheet read, for $||x||\gg 1/\sqrt{\sigma(\Lp)}$,
\be
\< h(0) h(x)\> = -\left(\frac{\Lp}{4\pi}\right)^2\log||x||.
\ee
However the vortex operator becomes
relevant at that point, and (anti)vortices start to proliferate. 

If one is approaching $\Lp^*$ from below, that is, 
from the `high temperature' phase of the XY model,
another universal property of the KT phase transition is the 
rapidly vanishing mass gap 
\be
  m \sim m_0\exp\left(-b\sqrt{\frac{K^*}{K^*-K}}\right)~.\la{eq:scaling}
\ee
$m_0$, $K^*$ and $b$ are non-universal quantities; but
given the function $\sigma(\Lp)$, this translates through Eq. 
\ref{eq:KK} into a 
robust prediction for the functional form of $m$ as a function of 
$\Lp$ close to the phase transition. ($\sigma(\Lp)$ is
defined by the asymptotic static force; we are assuming here
that $\sigma(\Lp)$ also governs the quantum fluctuations 
(see section 3).)
So the KT nature of the transition can in principle be tested,
but we note that in practice 
the range where the scaling~(\ref{eq:scaling})
holds is typically  very small~\cite{cardy}.

When a mass gap $m$ appears, the 
Casimir energy must decrease exponentially in $mL$
when the product is large. For a theory of free, neutral
massive bosons with Dirichlet boundary conditions, 
one can indeed show (see Appendix A) that
\be
E_C(L)= -\frac{1}{4} \sqrt{\frac{m}{\pi L}}~e^{-2mL}+
     O(e^{-4mL})  \la{eq:Ec_massive}
\ee
For the current theory, which contains vortices and 
anti-vortices, one expects twice this expression; 
and to obtain the corresponding expression for periodic
boundary conditions, one can apply the 
rule~\cite{ambjorn-wolfram}
\be
E_C(L)_{\rm periodic}~=~ 
 2 ~ E_C\left(\frac{L}{2}\right)_{\rm Dirichlet}. 
\ee
The Casimir energy is thus smaller than any power
of $1/L$, and the effective central charge vanishes.
\emph{So the phase transition of the effective theory 
driven by vortices explains
the disappearance of the L\"uscher 
correction as a transverse dimension is reduced in size.}
Although the vortices are not free, in the sine-Gordon 
model they correspond to solitons, 
which are the only one-particle states for 
$\beta^2>4\pi$, where $\beta^2\simeq\sigma(\Lp)\Lp^2$.
Their S-matrix 
is known exactly~\cite{zamolo}, and the Casimir energy
can be obtained from an exact non-linear integral 
equation~\cite{DdV}: in a periodic box, it reads
\be
E_C(L)=-\frac{2m}{\pi}K_1(mL) ~+~ \sqrt{\frac{\pi}{mL}}e^{-2mL}
\left[ \frac{1+\sqrt{2}}{2}+ O\left(\frac{1}{\sqrt{L}}  \right)  \right].
\ee
The leading term is  identical
with the free, complex, scalar field with the same 
periodic boundary conditions.
Thus this analytic form is universal, for a given mass gap
$m$.

\paragraph{The spread of the cross-over for a finite-length string}
Strictly speaking, the worldsheet statistical mechanics system
cannot have a phase transition for a finite-length string, 
instead the latter is generically smoothed out into a cross-over.
Decreasing $\Lp$ a little below $\Lp^*$
leads to a suppression of the $1/R$ term, which becomes
exponential in $m_0 R$ within
\be
\frac{\delta K}{K^*}= \frac{b^2}{\log^2 m_0R }~.
\ee
From then on, this transverse dimension is effectively
``switched off'' from the string dynamics. 
The argument above shows that the width in $K$ 
(and presumably also in $\Lp$) of the smoothed-out 
phase transition is only logarithmically suppressed in the 
string length, a distinctive signature of the KT nature of the
transition.

Since $m_0$ is expected to be of order $\sqrt{\sigma(\Lp)}$, 
what we just learnt has the practical consequence 
that $R$ must be exponentially
large in the correlation length of the underlying gauge theory, 
for this cross-over to be observed at all.
This might well explain why these effects
were not observed in the numerical studies~\cite{teper93,lucini01}
of spatial torelon masses in 2+1 dimensional gauge theories. 

\paragraph{The transition region for a finite-length string}
No matter how long the string is, for $\Lp$ sufficiently close
to $\Lp^*$ the mass gap becomes much smaller than $1/R$.
To get an idea of how the Casimir energy of the string then looks 
like, we compute in the appendix 
the leading corrections in the small $mR$ regime to the
conformal field theory value $-\pi/24R$ in the neutral 
free scalar field case:
\be
E_C=-\frac{\pi}{24R} ~+~ O(m) ~+~ O\left(m^2R\log(mR)\right).
\qquad(\sigma^{-1/2}(\Lp) \ll R \ll m^{-1})
\ee
More precisely, in terms of the static force, we have
\be
|F(\Lp,R)|=\sigma(\Lp)+\frac{\pi}{24R^2}+\frac{m^2}{4\pi}\log{mR}+\dots
\qquad(\sigma^{-1/2}(\Lp) \ll R \ll m^{-1})
\ee
It is effectively as if the string tension had a logarithmic dependence on
$R$ in this regime.

In the case of periodic boundary conditions, the leading term in $E_C$
would be $-\frac{\pi}{6R}$. This is also the outcome of an 
exact calculation in the sine-Gordon theory~\cite{DdV}.
Interestingly, in this regime the solitons and anti-solitons
do not lead to a doubling of the effective central charge;
on short distance scales, fluctuations are of the ``spin wave type''
dictated by the kinetic term of a \emph{real} scalar field.
The Casimir energy of the sine-Gordon model thus interpolates between
the Casimir energy of a free complex scalar field in the infrared
and a free real scalar field in the ultraviolet.
\subsection{Additional transverse dimensions \& applications}
So far we discussed the dynamics of a single transverse
dimension, in other words we were dealing with a three-dimensional
gauge theory. Now we return to the four-dimensional case,
where the two components of the worldsheet fluctuation
field $h$ are periodic, the case of relevance to standard
numerical simulations.

The simple but key point to note is that the two components are 
decoupled in the quadratic, renormalisable action~(\ref{eq:gauss_action})
responsible for the L\"uscher term. Therefore the universal 
contributions to the string energy levels just add up. 
Also the existence of topological configurations in one component 
is unaffected by the presence of the other component.
In general, the two XY-type models for the two components are 
coupled by `irrelevant' terms. This implies in particular 
that the energy of a vortex-antivortex `dipole' of size $L_{v\bar v}$
receives contributions of order $1/(\sigma(\Lp) L_{v\bar v}^2)$.
This however does not affect the Kosterlitz-Thouless criterion
for the occurrence of a phase transition driven by the vortices.
We therefore do not expect the KT nature of the
worldsheet phase transition to be affected by an additional 
fluctuation component.
Thus, if starting from the 4D theory, we \emph{successively}
reduce two dimensions, the latter get in turn `switched off' 
(from the point of view of contributing to the central charge) 
at two different KT phase transitions. If we reduce them
simultaneously, there presumably are two nearby transitions; 
this situation however deserves a more detailed investigation.

A convenient way to locate the worldsheet KT transition
in Monte-Carlo simulations of the lattice gauge theory
by working at a relatively small $L_1\equiv\Lp<\Lp^*$ 
is to measure the static quark potential off-axis, 
say at an angle $\alpha$: 
$\<P_4(0)P_4^*(R\cos\alpha \hat e_3+R\sin\alpha \hat e_1) \>$. 
The effective periodicity of the worldsheet fluctuation field
is then given by the substitution 
$\Lp\to \Lp(\alpha)\equiv\Lp/\cos\alpha$.
By measuring the static potential $V_\alpha(R)$, one can
in principle locate the $\alpha^*$ where the L\"uscher
term drops by a unit of central charge.
One can then check in an independent simulation 
that the transition is also seen when setting 
$\Lp=\Lp(\alpha^*)$ and measuring the static potential on-axis.
This would test that the transition is really driven
by the periodicity of the fluctuation field $h(z)$.

\section{Summary and outlook\la{sec:concl}}
We have studied an effective string theory for the 
flux-tube in SU($N$) gauge theories 
in three and four compact dimensions.
Let us summarise the outcome of our investigation of 
the Polyakov loop correlator when one transverse dimension
is periodic and of size $\Lp$.
The periodic component of the fluctuation field is described
by an action of the XY-model type. It follows that 
\bi
\item there is a phase transition of the worldsheet 
      field theory at $\Lp^*=O(\sigma^{-1/2})$ where the
      periodic component acquires a mass gap and hence the central
      charge drops by one unit;
\item it is a Kosterlitz-Thouless phase transition:
\item approaching $\Lp^*$ from below, the mass gap $m(\Lp)$ goes to zero
      in a universal way in terms of $K\equiv \sigma(\Lp)\Lp^2/4\pi^2$
      as given by Eq.~\ref{eq:scaling}. Once $\sigma(\Lp)$ is independently
       determined, 
      we have a prediction for the functional form of $m(\Lp)$. 
\item in the large $\Lp$ phase, there are (strongly suppressed) 
      corrections to the string
      energy levels associated with the periodicity;
      they come in powers of $1/R$ which vary continuously and increase 
      monotonously with $\Lp$; these corrections
      are exactly of order $1/R^3$ at $\Lp=\Lp^*$.
\ei

We conclude by mentioning some natural 
extensions of the investigations carried out in this paper.

\paragraph{1.} It is also possible to study the QCD string when one
transverse dimension is finite with Dirichlet boundary 
condition for the gauge field, as in the Schr\"odinger 
functional (see for instance~\cite{luscher_cargese}). 
The first question to address is what this boundary condition
translates into for the worldsheet degrees of freedom;
presumably it introduces a cutoff in the amplitude of the 
fluctuations. Unlike periodic boundary conditions, 
this cutoff will affect even the ``spin waves'' 
of the worldsheet.
It is known from the  discussion of the roughening 
phenomenon in lattice QCD that the bosonic string 
wave functional is Gaussian in the transverse coordinates
and that its width grows with the string length $R$ as~\cite{lmw}
\be
\delta^2(R)=\frac{1}{\pi\sigma}\log{\frac{R}{R_o}}.
\ee
One would thus expect that until $R$ is $O(e^{\sqrt{\sigma}\Lp})$,
the effect on the string spectrum 
is suppressed by $e^{-\Lp^2/\delta^2(R)}$.
At fixed $\Lp$, this offers a chance to verify the 
quantum-mechanical broadening of the string. 

\paragraph{2.} In string theory the duality transformation 
$\Lp\leftrightarrow \frac{2\pi}{\sigma \Lp}$ plays an important
role. It is clear that a self-dual point at $\Lp\simeq 0.18$fm 
does not exist in the \sun theory at finite $N$.
From the point of view taken in this paper, 
it is because the string tension and the effective string action
varies with $\Lp$, unlike a fundamental string theory.

\paragraph{3.} The fact that the QCD flux-tube,
when probed at energies well below the scale set by its tension,
behaves like a bosonic string
living in the same number of dimensions as the gauge theory,
is highly remarkable. One might call this the ``weak gauge-string
equivalence''~\cite{lw04}. Besides this line of investigation, there
is the long-standing speculation that the gauge theory itself
has an equivalent formulation in string theory (the ``strong
gauge-string equivalence''~\cite{polyakov2}). 
Since the gauge theory is renormalisable, it is a stand-alone theory
which describes phenomena on all scales. 
Therefore the equivalent string formulation must also
make sense at all energy scales, which means that it must be
a fundamental string theory. All known fundamental string theories
live in a higher number of dimensions.
                                                                             
It is natural to ask what the relation is between the
effective four-dimensional string, whose properties can be
established through the weak gauge-string correspondence,
and the hypothetical, presumably higher-dimensional string theory
which behaves like the gauge theory on all scales. One would naively
expect the former to arise as the low-energy effective string
obtained by dimensionally reducing the degrees of freedom of
the latter, fundamental string theory. 
If such a relation exists, the worldsheet phase transition 
of the effective string described 
in this paper must also affect the fundamental string
dynamics, and hence should be reflected in other physical observables
of the gauge theory. The
gauge theory of course has the `deconfining' phase transition at
$\Lp\doteq 1/T_c=O(\sigma^{-1/2})$, and it is clearly interesting to see whether
$\Lp^*=1/T_c$. This can be achieved straightforwardly by lattice simulations.
Since dimensional reduction in the gauge theory does not 
work all the way down to $T_c$, 
there is no \emph{a priori} reason to expect $\Lp^*=1/T_c$: 
such a coincidence would provide evidence for a relation
of the effective QCD string to a fundamental string.   

The low-energy
constants determined in the weak gauge-string correspondence
also provide valuable information on the `mother' string theory,
and the geometry of the space which leads to its dimensional reduction.
Their dependence on the number of colours is of particular 
interest~\cite{meyer04}, since the equivalence to a string theory
is expected to be simpler at large $N$.

\paragraph{4.} We note that the Kosterlitz-Thouless 
worldsheet transition was discussed 
in the context of the bosonic string, 
as a fundamental theory in 26 dimensions,
by Sathiapalan~\cite{sathiapalan}. 
The fact that the $O(2)$ nonlinear $\sigma$-model has
no fix points for $\kappa<\kappa^*$ implies that it cannot
be the worldsheet action of a \emph{fundamental} string theory.
The author concludes that in this regime, 
the string can no longer be considered to be
in a flat background, because the vortex condensate 
contributes to the stress-energy tensor and hence to
the space-time curvature.
In the case of the QCD string, we have a fixed cutoff, 
and we have shown that the net effect is a change of
the central charge, as measured 
through the Casimir energy of the string: one 
component of the worldsheet field effectively 
acquires a mass. Thus an attractive scenario for 
a fundamental string is the following. 
When probed at low energies $E\ll T^*$, the Casimir energy
appears to be that of a string in lower dimensions, due
to the curvature of the target space-time which gives an
effective mass to some components of the worldsheet field.
But at higher energies, the curvature is irrelevant and 
the ``hidden'' components can be excited:  the
original conformal invariance is restored in this regime.
In this way, it is plausible that there exists a fundamental
string theory which at low-energies describes the QCD string.
\section*{Acknowledgements}
I thank Janos Balog, Rainer Sommer and Peter Weisz
for a critical reading of the manuscript and the ensuing
discussions which helped me to clarify several issues.
\appendix
\input{integral.tex}

\end{document}

%% file: pic1-insert.tex
\begin{figure}
\hspace{4cm}
\vspace{1.8cm}
\begin{minipage}[l]{7cm}
\begin{picture}(0,0)(90,160)
\put(0,0){\line(1,0){200}}
\put(0,90){\line(1,0){200}}
\put(0,180){\line(1,0){200}}
\put(60,72){\circle*{4}}
\put(53,66){${\rm v}$}
\put(120,108){\circle*{4}}
\put(121,108){$\bar{\rm v}$}

\put(-25,4){$t=-\infty$}
\put(-25,94){$t=0$}
\put(-25,184){$t=+\infty$}

\put(0,-10){$0$}
\put(196,-10){$R$}

 \thicklines{
 \put(0,0){\line(0,1){180}}
 \put(204,0){\line(0,1){180}}

 \put(0,0){\vector(1,0){14}}
 \put(30,0){\vector(3,1){13}}
 \put(50,0){\vector(3,-1){13}}
 \put(90,0){\vector(3,-1){13}}
 \put(120,0){\vector(3,1){14}}
 \put(150,0){\vector(3,1){14}}
 \put(190,0){\vector(1,0){14}}

 \put(0,90){\vector(1,0){14}}
 \put(30,90){\vector(3,-1){12}}
 \put(50,90){\vector(1,-2){5}}
 \put(90,90){\vector(-1,0){14}}
 \put(120,90){\vector(0,1){14}}
 \put(150,90){\vector(1,1){10}}
 \put(190,90){\vector(1,0){14}}

 \put(0,180){\vector(1,0){14}}
 \put(30,180){\vector(3,-1){13}}
 \put(50,180){\vector(3,-1){13}}
 \put(90,180){\vector(3,1){13}}
 \put(120,180){\vector(3,1){14}}
 \put(150,180){\vector(3,-1){14}}
 \put(190,180){\vector(1,0){14}}

 }
\end{picture}
\hspace{8cm}
\begin{picture}(0,0)(50,190)
\put(-45,29){{\huge$\Leftrightarrow$}}
\put(-45,119){{\huge$\Leftrightarrow$}}
\put(-45,209){{\huge$\Leftrightarrow$}}
 \put(0,0){\line(1,0){50}}
 \put(0,20){\line(1,0){50}}
 \put(0,40){\line(1,0){50}}
 \put(0,60){\line(1,0){50}}

 \put(0,90){\line(1,0){50}}
 \put(0,110){\line(1,0){50}}
 \put(0,130){\line(1,0){50}}
 \put(0,150){\line(1,0){50}}

 \put(0,180){\line(1,0){50}}
 \put(0,200){\line(1,0){50}}
 \put(0,220){\line(1,0){50}}
 \put(0,240){\line(1,0){50}}

\put(5,25){\circle*{4}}
\put(-2,29){${\rm q}$}
\put(45,25){\circle*{4}}
\put(46,29){$\bar{\rm q}$}
\put(45,45){\circle*{4}}
\put(45,5){\circle*{4}}

\put(5,115){\circle*{4}}
\put(-2,119){${\rm q}$}
\put(45,95){\circle*{4}}
\put(46,99){$\bar{\rm q}$}
\put(45,135){\circle*{4}}
\put(45,115){\circle*{4}}

\put(5,205){\circle*{4}}
\put(-2,209){${\rm q}$}
\put(45,205){\circle*{4}}
\put(46,209){$\bar{\rm q}$}
\put(45,185){\circle*{4}}
\put(45,225){\circle*{4}}

 \thicklines{
 \put(5,25){\line(1,0){40}}
 \put(5,115){\line(2,-1){40}}
 \put(5,205){\line(1,0){40}}

\put(60,29){$t=-\infty$}
\put(60,119){$t=0$}
\put(60,209){$t=+\infty$}
}
\end{picture}
\end{minipage}
\vspace{6.0cm}
\caption{Physical interpretation of the vortices: the left-hand side,
where $\theta$ is represented as a unit vector $(\cos\theta,\sin\theta)$,
illustrates that the appearance of a vortex at time $t$ `followed' by an 
antivortex at time $\bar t$ induces one unit of clockwise winding 
of the variable $\theta(t_s,x)$ as  $x$ varies from $0$ to $R$
on a timeslice at time $t_s$ located between $t$ and $\bar t$.
Right, the corresponding interpretation for the  
string on the covering space of the torus: the string has non-trivial 
winding number along the transverse dimension in the interval
$[t,\bar t]$ and trivial winding number elsewhere.
} 
\label{fig:winding}
\end{figure}
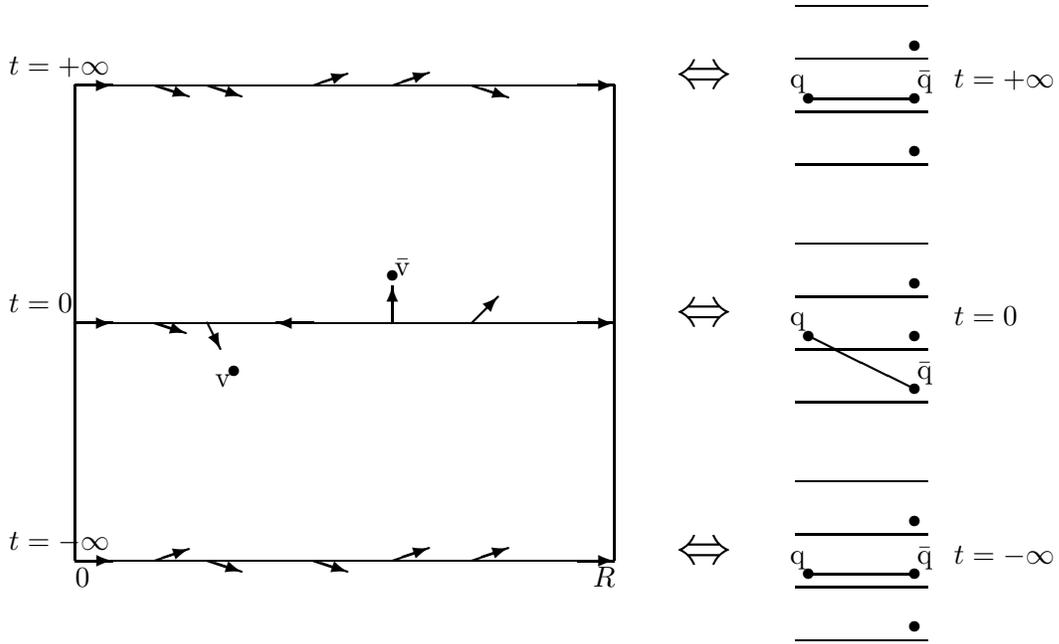

%% file: integral.tex
\section{The Casimir effect for a massive scalar field}
The Casimir energy for a free massive scalar field in $d$ space dimensions
with Dirichlet boundary condition in one dimension of size $R$
reads~\cite{ambjorn-wolfram}
\be
E_C(m,R)=\frac{1}{(2\sqrt{\pi}R)^d~\Gamma\left(\frac{d}{2}\right)}~
    \int_0^\infty dt~ t^{d-1}~\log\left(1-e^{-2\sqrt{t^2+m^2R^2}}\right)
\la{eq:Ec1}
\ee
This is the energy per unit of transverse volume. A term proportional
to $R$ and a term proportional to $m$
have already been discarded at this stage~\cite{ambjorn-wolfram}.
By expanding the logarithm $\log(z)$ around $z=1$ and integrating 
term by term, one obtains $E_C$ as a series of modified Bessel 
functions:
\be
E_C(m,R)=\frac{-2}{R^d}
\left(\frac{mR}{4\pi}\right)^{\left(\frac{d+1}{2}\right)}
\sum_{n=1}^\infty 
\frac{K_{\frac{d+1}{2}}(2mRn)}{n^{\left(\frac{d+1}{2}\right)}}.
\la{eq:Ec2}
\ee
Given the asymptotic large-argument behaviour of the Bessel functions,
\[K_\nu(z)\sim \sqrt{\frac{\pi}{2z}}e^{-z},\qquad z\to\infty,~\forall\nu  \]
this expression of $E_C$ immediately shows that the latter vanishes
exponentially fast in $mR$ once this product is large:
\be
E_C(mR\to\infty,R)\sim - \frac{1}{2R^d}\left(\frac{mR}{4\pi}\right)^{d/2}
                         e^{-2mR}.
\ee

On the other hand, when 
\[\epsilon \equiv mR\ll 1, \]
one naturally expects to recover the Casimir energy of 
a massless scalar field~\cite{ambjorn-wolfram}
\be
E_C(m=0,R) =- \frac{1}{(4\pi)^{\frac{d+1}{2}}}~\frac{1}{R^d}
~\Gamma\left(\frac{d+1}{2}\right) \zeta(d+1).
\ee
This is easily checked to be the case, either by using the 
small-argument expansion of the Bessel functions,
\be
 K_\nu(z)\sim \frac{\Gamma(\nu)}{2}~\left(\frac{z}{2}\right)^{-\nu},  
\la{eq:Knu_small}
\ee
or by contour integration techniques of~(\ref{eq:Ec1}) (with $mR=0$).
The leading corrections to this was given in~\cite{ambjorn-wolfram}
for $d\geq3$:
\be
E_C(m,R)=E_C(m=0,R)~+~ \frac{1}{(4\pi)^{\frac{d+1}{2}}}~\frac{1}{R^d}
\Gamma\left(\frac{d-1}{2}\right)\zeta(d-1)(mR)^2 ~+~
\dots
\ee
For $d=2$, the leading correction is easily computed by
the change of variables $t^2=s^2-\epsilon^2$ in (\ref{eq:Ec1}):
\be
E_C(m,R)=E_C(m=0,R) -  \frac{m^2}{8\pi} \left[\log\left(2mR\right)-1\right]
+O(mR)^3, \qquad d=2.
\ee

We are interested in the Casimir energy of a string, so that $d=1$ and,
using \eq\ref{eq:Knu_small} and $\zeta(2)=\frac{\pi^2}{6}$, 
we have the well-known result~\cite{luscher81}
\be
E_C(m=0,R)=-\frac{\pi}{24R}.
\ee
We now want to compute the leading corrections in $\epsilon$ to this
result, which corresponds to the conformal theory. 
This is a less straightforward calculation, because one cannot 
simply differentiate (with respect to $\epsilon$)
the integrand in the first form~(\ref{eq:Ec1}), or the terms of the series
in the second form~(\ref{eq:Ec2}), 
because in both cases the result would diverge.
We start off from~(\ref{eq:Ec2}).
The key point to note is that as $\epsilon\to0$, the series
samples the Bessel function in very small steps, so that in can
be approximated by an integral:
In general, for a smooth function $g(x)$, one would have
\be
\epsilon \sum_{n\geq1} g(\epsilon n) = 
\int_{\epsilon/2}^\infty g(x) dx + O(\epsilon^2)
\ee
This is however not directly applicable, because the $K_1(x)$
function is singular at the origin. We therefore differentiate
$E_C$ with respect to $\epsilon$ at fixed $R$:
\ba
-2\pi R \frac{\partial E_C(\epsilon,R)}{\partial\epsilon} &=&
  \sum_{n\geq1} \frac{K_1(2\epsilon n)}{n}~+~2\epsilon K_1'(2\epsilon n)\\
  &=& -2\epsilon \sum_{n\geq1} K_0(2\epsilon n) 
\ea
where $K_1'$ is the derivative of $K_1$ with respect to its argument.
We have used the identity $K_1'+K_1/z=-K_0$.
We now use the formula, valid for an analytic function,
\be
2\epsilon \sum_{n\geq 1} f(2\epsilon n)=
\int_\epsilon^\infty f(x)dx - 2\epsilon \sum_{n\geq1}
\sum_{k\geq1} \frac{f^{(2k)}(2n\epsilon)}{(2k+1)!}\epsilon^{2k}.\la{eq:ee}
\ee
Now, with $f=K_o$, 
\be
\int_\epsilon^\infty K_o(x)dx = 
\int_0^\infty K_o(x)dx-\int_0^\epsilon K_o(x)dx
=\frac{\pi}{2}+\epsilon \left[\log\epsilon -1-\log2+\gamma   \right]
 + O(\epsilon^3\log \epsilon).
\ee
The remaining sums in~(\ref{eq:ee}) can be evaluated in leading 
order in $\epsilon$ by using the leading 
small-argument behaviour of $K_o(\epsilon)=-\log\epsilon + O(1)$:
\be
\sum_{n\geq1} K_o^{(2k)}(2n\epsilon)\epsilon^{2k}
= \frac{(2k-1)!\zeta(2k)}{4^k} + O(\epsilon^2\log\epsilon)
\ee
Gathering the pieces and performing a few elementary integrals, we get
\ba
E_C(m,R)&=&E_C(0,R)+\int_0^{\epsilon} 
\frac{\partial E_C(\epsilon',R)}{\partial \epsilon'}d\epsilon'\nn
&=&-\frac{\pi}{24R}+ \frac{m}{4} + \frac{m^2R}{4\pi}
\left[ \log\left(\frac{mR}{2}\right)+\gamma-\frac{3}{2}-\sum_{k\geq1}
\frac{\zeta(2k)}{4^k k(2k+1)}   \right] \nn
 && + O(m^4R^3\log mR)
\ea
It is clear from the results above that the introduction 
of a small mass for the scalar field has a stronger effect
in lower space dimensions. This is natural since infrared 
effects tend to be stronger in lower-dimensional field theories.
The term proportional to $m^2R\log mR$ is the first term of
interest, in the sense that it is the leading observable effect
in a system with fixed $m$ and varying $R$.
It expresses the fact that the bulk free energy of the system
has a logarithmic dependence on its size, in the regime $mR\leq 1$.

The cross-over between the small $mR$ and large $mR$ regimes is smooth.
The area $-R\int_0^\infty d\epsilon E_C(\epsilon,R)$
is easily computed and amounts to $\zeta(3)/16$.